\documentclass[10pt]{article}
\usepackage{amsmath}
\usepackage{amssymb,amsthm}
\usepackage{graphicx}
\usepackage{cite}
\usepackage{pifont,bbding}
\usepackage{color}
\usepackage[usenames,dvipsnames]{xcolor}

\usepackage[labelfont=bf,labelsep=period,justification=raggedright]{caption}
\makeatletter
\renewcommand{\@biblabel}[1]{\quad#1.}
\makeatother

\date{}

\newtheorem{lemma}{Lemma}

\newcommand{\ignore}[1]{}

\begin{document}

\begin{flushleft}
{\Large
\textbf{A combinatorial model of malware diffusion via Bluetooth connections}
}
\\
Stefano Merler,
Giuseppe Jurman$^\ast$ 
\\
Fondazione Bruno Kessler, Trento, Italy
\\
$\ast$ E-mail: jurman@fbk.eu
\end{flushleft}

\section*{Abstract}
We outline here the mathematical expression of a diffusion model for cellphones malware transmitted through Bluetooth channels. 
In particular, we provide the deterministic formula underlying the proposed infection model, in its equivalent recursive (simple but computationally heavy) and closed form (more complex but efficiently computable) expression.

\section*{Introduction}
The spreading of malware, \textit{i.e.}, malicious self-replicating codes, has rapidly grown in the last few years, becoming a substantial threat to the wireless devices, and mobile (smart)phones represent nowadays the most appetible present and future target.
Papers studying the problem from both theoretical and technical points of view already appeared in literature since 2005 \cite{wang09understanding,nguyen09novel,zanero09wireless,guanhua09modeling,shuai09modeling,khouzani11optimal,ghallali11mobile,lapolla12survey,ghallali12designing}, and nowadays a number of different approaches to modeling the virus diffusion are already available to the community.
With the present work we want to contribute to this topic by proposing a more accurate model for the spread of a malware through the Bluetooth channel, providing both a recursive and a combinatorial equivalent deterministic formulation of the described solution.

\section*{The model}
The dynamics of the proposed model is the following: at a certain time $\tau$, a number $I$ of infected mobiles $b_1,\ldots,b_I$ come in contact with a number $S$ of clean (non-infected) cellphones $w_1,\ldots,w_S$; hereafter we will denote this configuration as $(I,S)$.  

All $S+I$ telephones are in the Bluetooth transmission range of each other and they all have their Bluetooth device on. 
Each infected mobile tries to establish a connection with another device, clearly not knowing whether it is trying to pair to a clean or to an infected phone. 
All these connections are established instantaneously at time $\tau$. 
However, for the sake of simplicity we assume that the infected mobiles establish connections following a given sequence, starting from $b_1$ down to $b_I$. 
In other words, $b_1$ is the first to try to establish a connection, $b_I$ is the last one.
Moreover, each connection is chosen uniformly at random among all possible available choices.
Connections between infected and clean mobiles deterministically result in infection transmission: when a clean mobile gets paired to an infected one, it becomes infected. 
All these events occur in the time interval $[\tau,\tau+\Delta\tau]$, where $\Delta\tau$ is the minimal time allowing all infected mobiles to establish a connection and eventually transmit the virus: in practice, it may be considered of the order of a few tens of seconds. 
We assume that in this time interval clean cellphones do not try to establish any connections, \textit{e.g.}, for non-malware purposes. 
We also assume that in this time interval no other mobile enters the Bluetooth transmission range of the $S+I$ mobiles and, when a connection between two mobiles is established, the two mobiles remain connected for the whole time interval. 
Basically, we are assuming that the initial configuration $(I,S)$ is given and it does not change in the time interval $[\tau,\tau+\Delta\tau]$. 
Note that, given the definition of $\Delta\tau$, new infections do not result in configuration changes in the time interval $[\tau,\tau+\Delta\tau]$. 

All the aforementioned assumptions are reasonably realistic, due to the very short time-scale considered.

The task here is to discover the probability that, in this situation, a given clean mobile gets paired to an infected one, and thus it becomes itself infected.

Summarizing, the setup and the constraints of the model are the following:
\begin{description}
\item[Setup] $I$ infected mobiles $b_1,\ldots,b_I$ and $S$ clean mobiles $w_1,\ldots,w_S$ are in a room (\textit{i.e.}, in the Bluetooth transmission range of each other).
\item[Dynamics] Starting from $b_1$ down to $b_I$, each infected mobile tries to connect with a yet unconnected device, regardless of whether it is infected or not.
\item[Constraint \#1] Since the connection channel is Bluetooth, once a connection between two mobiles is established, these two devices become unavailable to further connection, or, in other words, each device can have at most one connection to another cellphone.
\item[Constraint \#2] For each $t=1,\ldots,I$, when it is $b_t$'s turn to choose, $b_t$ must connect to one of the still available devices, if any.
\end{description}

\begin{figure}[!htp]
\begin{center}
\includegraphics[width=0.9\textwidth]{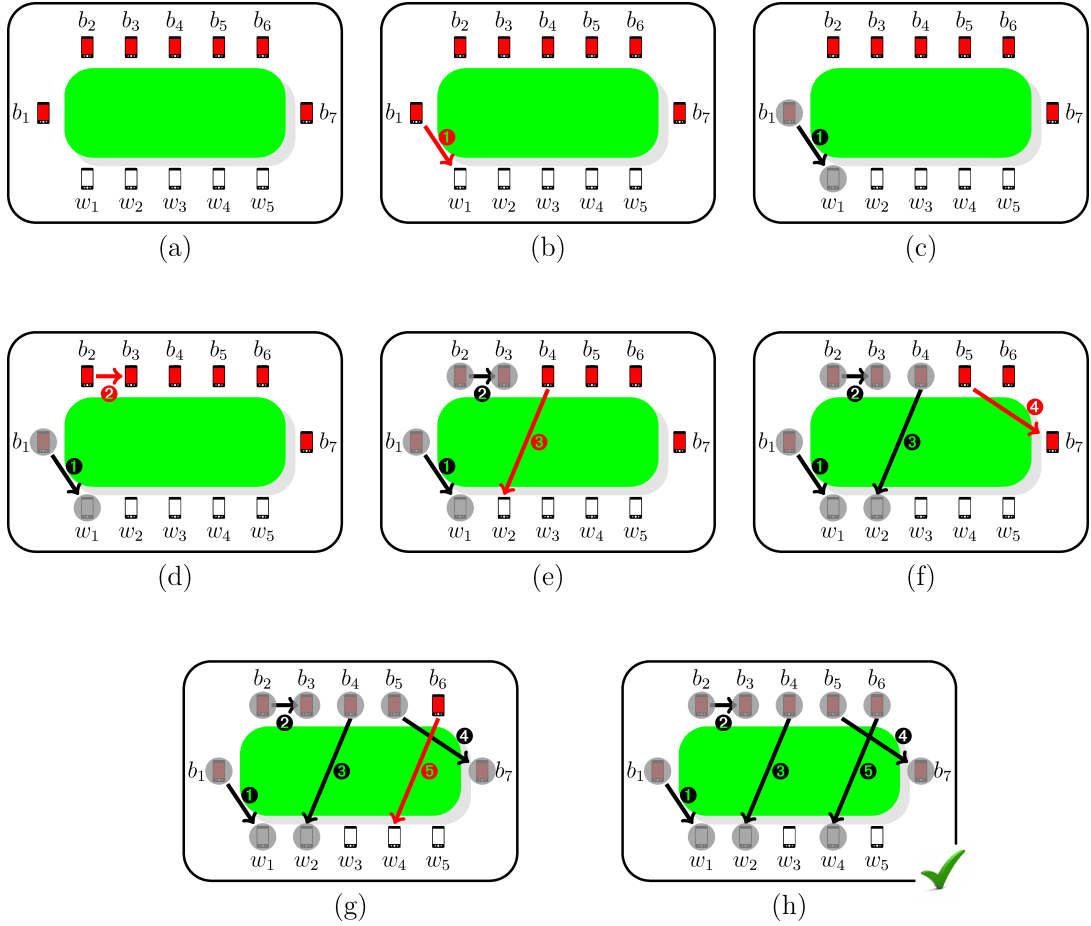}
\end{center}
\caption{
\textbf{An example of model dynamics starting from the initial configuration (7,5). In red, the pairing that it is established at each step.} 
(a) At time $\tau$, $I=7$ infected mobile phones $b_1,\ldots,b_7$ and $S=5$ clean mobiles $w_1,\ldots,w_5$ are all within their mutual Bluetooth connection range.
(b) $b_1$ chooses a mobile among $b_2,\ldots,b_7,w_1,\ldots,w_5$; it chooses $w_1$ establishing connection \ding{202}.
(c) Now it $b_2$'s turn to choose, and $b_1$ and $w_1$ are not available anymore for pairing (marked by a grey circle \textcolor{gray}{\CircleSolid}).
(d) $b_2$ connects to $b_3$ through pairing \ding{203}.
(e) The two mobiles $b_2$ and $b_3$ become unavailable for pairing, too and the next infected mobile in line $b_4$ pairs to $w_2$ via \ding{204}.
(f) Only $b_6, b_7$ and $w_3,w_4,w_5$ remain available for pairing with $b_5$, which chooses $b_7$ (connection \ding{205}).
(g) Now the last mobile $b_6$ must connect to the remaining unpaired clean phones $w_3,w_4,w_5$: it chooses $w_4$ creating pairing \ding{206}.
(h) There are no more unpaired infected mobiles: the process ends at time $\tau+\Delta\tau$.
}
\label{fig:generic_example}
\end{figure}

Let us consider the generic configuration $(I,S)$ with $I$ unpaired infected mobiles $b_1,\ldots,b_I$ and $S$ unpaired clean mobiles $w_1,\ldots,w_S$. 
According to the setup, the first mobile establishing a connection is $b_1$. 
In Fig.~\ref{fig:generic_example} a possible evolution is displayed starting from an initial configuration with $I=7$ infected and $S=5$ clean mobiles, together with an explanatory description of the occuring dynamics.

\begin{figure}[!htp]
\begin{center}
\includegraphics[width=0.9\textwidth]{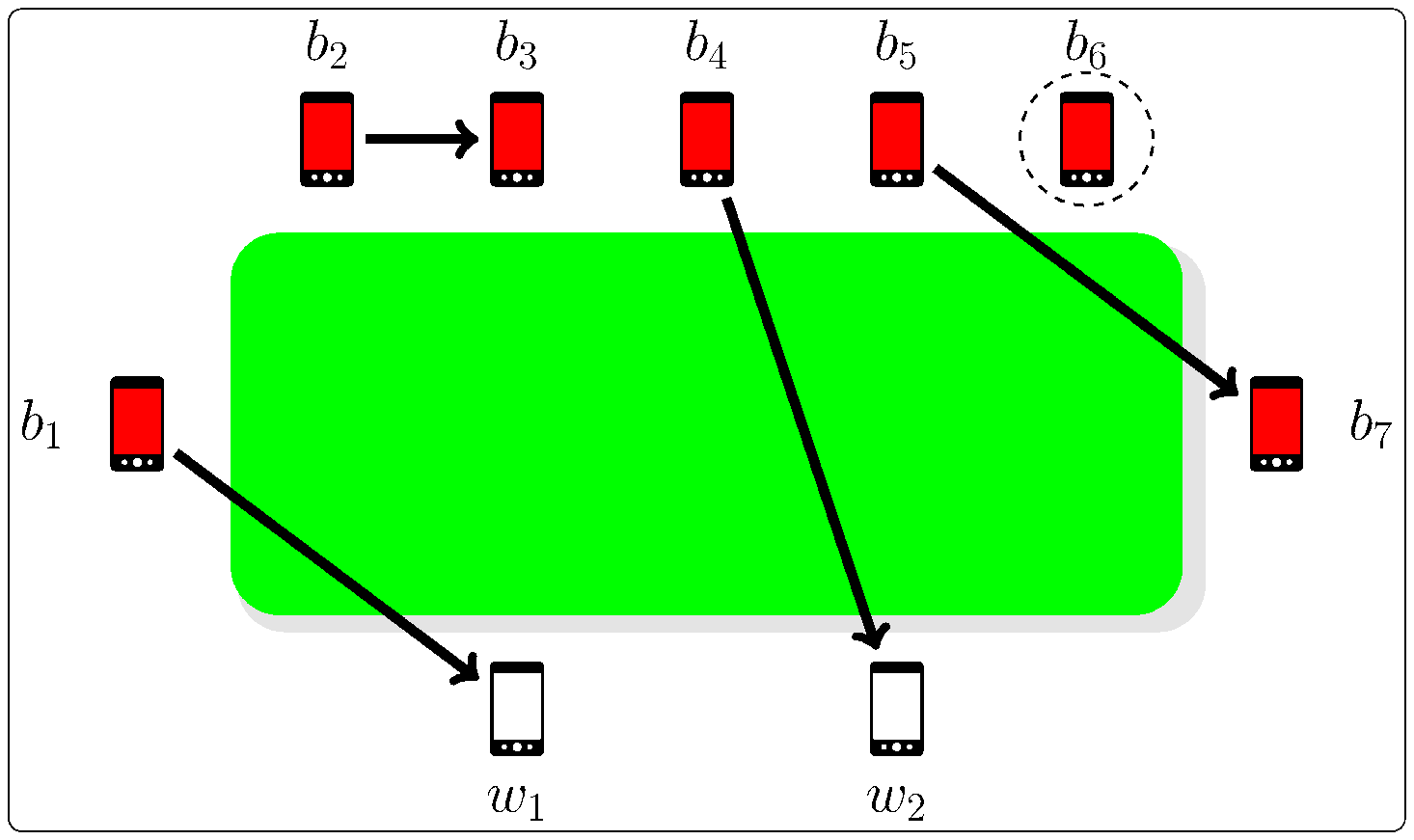}
\end{center}
\caption{
\textbf{An example of the $\dag$ situation.} Starting from the initial configuration $(7,2)$, $b_1$ infects the clean mobile $w_1$, $b_2$ pairs to $b_3$, $b_4$ infects $w_2$ and, finally, $b_5$ pairs to $b_7$. Here the process ends, because there are no more mobiles available for pairing to $b_6$ which remains unconnected.
} 
\label{fig:dag_case_example}
\end{figure}

Due to the described dynamics, all the infected mobiles succeed in paring, with the exception of at most one $b_z$, which can remain unpaired if there are no more available mobiles.
This case can only happen when there are more infected mobiles than clean ones, their sum is odd and all the clean mobiles get paired: 
\begin{equation*}\tag{\dag}
\left\{
\begin{split}
I &> S \\
I+S &\in 2\mathbb{Z}+1 \\ 
j &= \frac{I-S-1}{2}\ ,
\end{split}
\right.
\end{equation*}
where $j$ is the number of pairings between two infected mobiles.
Henceforth, the last choosing infected mobile $b_z$ cannot find any available device to pair to. 
In what follows, we will refer to this case as the case $\dag$; an example of this situation in the initial configuration $(7,2)$ is shown in Fig.~\ref{fig:dag_case_example}.

The model is completely described by computing the probability $P(I,S)$ that a certain clean mobile, for instance $w_1$, gets infected in the time interval $[\tau,\tau+\Delta \tau]$. 

Although $P(I,S)$ could be stochastically approximated by running repeated simulations, in the following Sections we will derive two equivalent exact (deterministic) formul\ae\ for $P(I,S)$ in the aforementioned setup.
The former is a simple recursive expression, which follows straightforwardly from the model dynamics, while the latter is its corresponding closed form (thus with no recursion involved), which has a more complex expression and it heavily relies on combinatorics.
Other than their alternative mathematical nature, the two formul\ae\  show different behaviours also from a computational point of view, as discussed in a dedicated Section.

\section*{The recursive formula}
Recursively, the probability $P(I,S)$ of a given susceptible mobile $w_t$ to get infected starting from a given initial configuration $(I,S)$ can be written by the following expression:
\begin{equation}
\boxed{
\left\{
\begin{split}
   P(I,S) &= {\frac{1}{I+S-1}} + {\frac{S-1}{I+S-1} P(I-1,S-1)} + {\frac{I-1}{I+S-1} P(I-2,S)}\\
   P(0,S) &= 0\\
   P(I,0) &= 0\\
   P(1,S) &= \frac{1}{S}\ .
\end{split}
\right.
}
 \label{eq:recursive}
\end{equation}
where the trivial conditions $P(0,S)=0$, $P(I,0)=0$ and $P(1,S)=1/S$ initialize the recursion, thus covering all possible cases.

Since all clean mobiles share the same probability $P(I,S)$ of getting infected, without loss of generality we may assume $w_t=w_1$.
The three terms $\frac{1}{I+S-1}$, $\frac{S-1}{I+S-1} P(I-1,S-1)$, and $\frac{I-1}{I+S-1} P(I-2,S)$ contributing to the general case of $P(I,S)$ come from the three mutually exclusive cases which can occur starting from the initial configuration $(I,S)$:
\begin{enumerate}
\item $b_1$ establishes a pairing with $w_1$. In this case $w_1$ gets infected and this event occurs with probability $\frac{1}{I+S-1}$.
\item $b_1$ establishes a pairing with one of the other $S-1$ clean mobiles $w_2,\ldots,w_S$. 
This event occurs with probability $(S-1)\cdot\frac{1}{I+S-1}$ and of course $w_1$ does not get infected by $b_1$. 
However, $w_1$ may be infected later by the remaining $I-1$ available infected phones (with only $S-1$ clean mobiles still available, because one clean mobile has been infected by $b_1$), thus falling back to a $(I-1,S-1)$ configuration.
\item $b_1$ establishes a pairing with one of the other $I-1$ unpaired infected mobiles $b_2,\ldots,b_I$.  This event occurs with probability $(I-1)\cdot\frac{1}{I+S-1}$ and of course $w_1$ does not get infected by $b_1$. 
However, similarly to the previous situation, $w_1$ may be infected later by the remaining $I-2$ unpaired infected phones, thus falling back to a $(I-2,S)$ configuration.
\end{enumerate}

\begin{figure}[!htp]
\begin{center}
\includegraphics[width=0.9\textwidth]{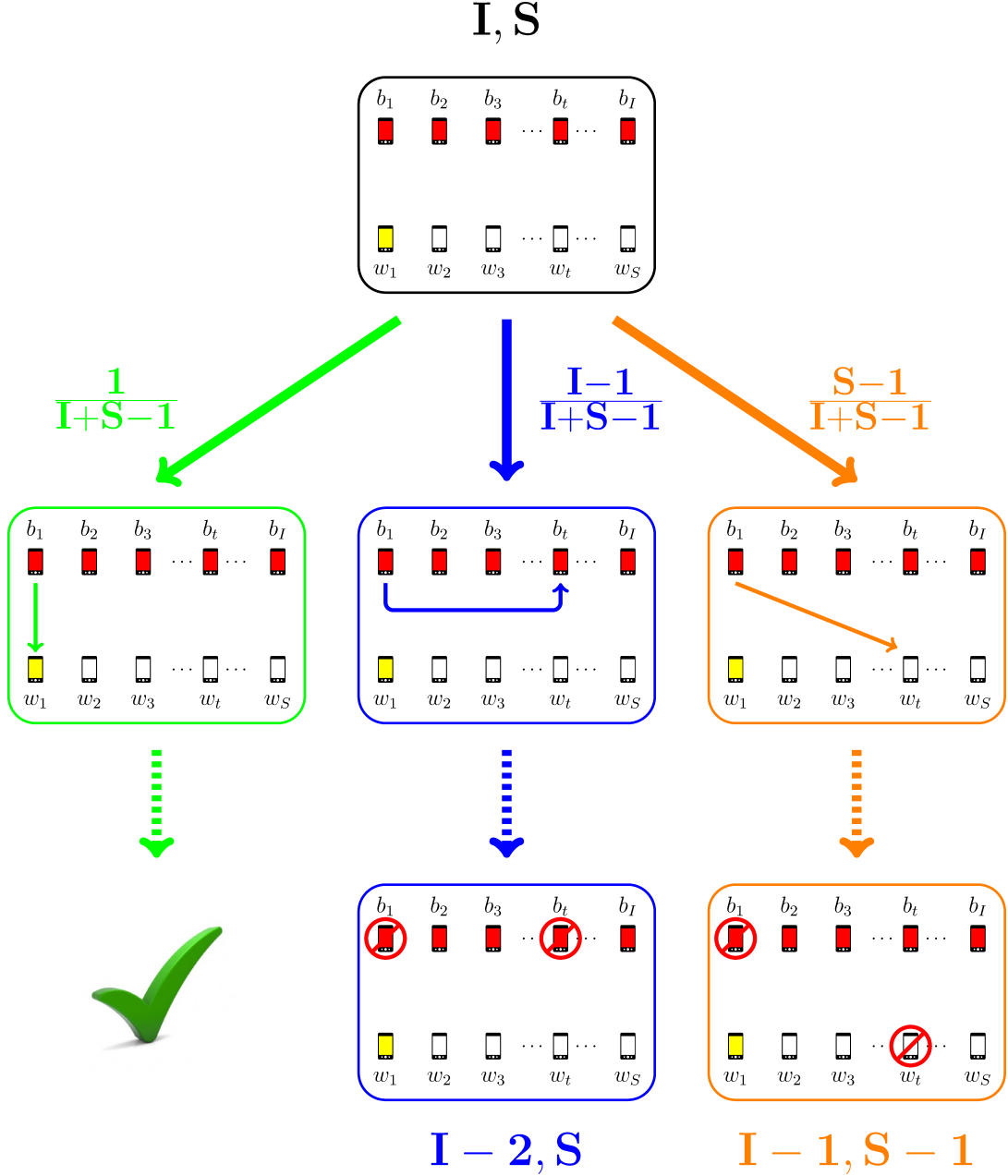}
\end{center}
\caption{
\textbf{Construction of the general case of the recursive formula Eq.~\ref{eq:recursive}} Starting from the initial configuration $(I,S)$, we want to compute the probability $P(I,S)$ that a clean mobile ($w_1$ without loss of generality) gets infected in the proposed model. 
At time $\tau$, the first infected mobile $b_1$ tries to establish a pairing, and only one of the three following alternatives can occur. 
In green, the case when $b_1$ immediately infects $w_1$ (with probability $\frac{1}{I+S-1}$) and we are done.
In blue, the case when $b_1$ pairs to one of the remaining another $I-1$ infected mobiles $b_t$ with probability $\frac{I-1}{I+S-1}$; then $b_1$ and $b_t$ becomes unavailable for pairing with the following choosing mobile $b_2$, and we are moved into the case of computing the probability that $w_1$ gets infected when there are $I-2$ unlinked infected mobiles and $S$ clean ones, \textit{i.e.}, $P(I-2,S)$.
Finally, in orange, the case when $b_1$ pairs to one of the other $S-1$ clean mobiles $w_t$ (with $w_t\not=w_1$) with probability $\frac{S-1}{I+S-1}$; then $b_1$ and $w_t$ becomes unavailable for pairing with the following choosing mobile $b_2$, and we are moved into the case of computing the probability that $w_1$ gets infected when there are $I-1$ unlinked infected mobiles and $S-1$ unlinked clean ones, \textit{i.e.}, $P(I-1,S-1)$.
The general case $P(I,S)={\frac{1}{I+S-1}} + {\frac{S-1}{I+S-1} P(I-1,S-1)} + {\frac{I-1}{I+S-1} P(I-2,S)}$ is obtained by summing the contributions of all three alternative cases described above.
}
\label{fig:recursive}
\end{figure}

A worked out example illustrating the construction of Eq.~\ref{eq:recursive} is shown in Fig.~\ref{fig:recursive}.

The formula in Eq.~\ref{eq:recursive} for $P(I,S)$ relies on a recursive equation of second order with non constant coefficients, for which no general method is known to derive the corresponding non-recursive (closed) expression. 
Moreover, as detailed in a later Section, calculating $P(I,S)$ by using Eq.~\ref{eq:recursive} is computationally heavy.
However, we will obtain the equivalent time-saving closed form solution in the next Section using combinatorial arguments.

\section*{The combinatorial formula}
\begin{figure}[!htp]
\begin{center}
\includegraphics[width=0.9\textwidth]{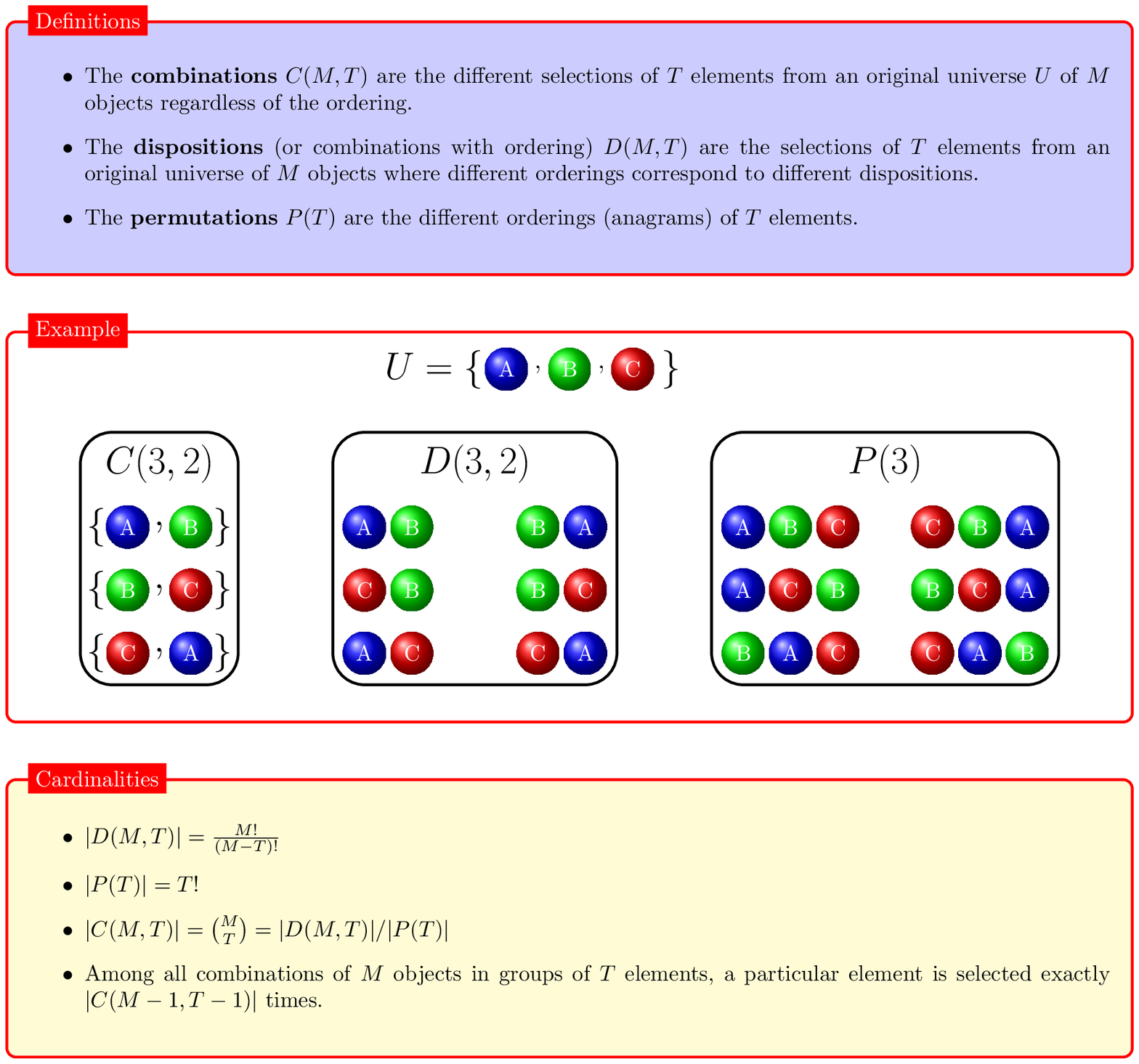}
\end{center}
\caption{\textbf{Basic definitions, examples and facts on dispositions, combinations and permutations.}}
\label{fig:combinatorics}
\end{figure}

To construct the explicit formula equivalent to Eq.~\ref{eq:recursive}, we need to employ a few combinatorial considerations.
The key observation is that we can count all wirings (lists of pairings) that can occur at the end of the pairing process.
Clearly, the fact that there is an order in setting up the connections between the mobiles heavily influences the probability that a given wiring can occur: in particular, this probability depends on the number $j$ of pairings between infected mobiles (bb-pairings, for short).
As background material, we recall some definitions and results from combinatorics in the box in Fig.~\ref{fig:combinatorics}, together with the two following functions:
\begin{itemize}
\item the Heaviside step function
\begin{displaymath}
H(x) = 
\begin{cases}
1 & \textrm{for $x\geq 0$} \\
0 & \textrm{for $x<0$}\ ;
\end{cases}
\end{displaymath}
\item the Kronecker delta function
\begin{displaymath}
\delta(x) = 
\begin{cases}
1 & \textrm{for $x=0$} \\
0 & \textrm{for $x\not =0$}\ ,
\end{cases}
\end{displaymath}
\end{itemize}

As an example, the following indicator function can be written in the two equivalent formulations:
\begin{displaymath}
\begin{split}
f(I,S,j) &= 
\begin{cases}
1 & \textrm{in the $\dag$ case}\\
0 & \textrm{otherwise}
\end{cases}\\
&= H(I-S-1)\delta((I+S+1) \bmod 2)\delta(2j-I+S+1)\ ,
\end{split}
\end{displaymath}
where $\mod$ is the Euclidean remainder function, so $x\bmod 2$ is zero for even $x$ and one for odd $x$.

Suppose now we are starting from an initial configuration $(I,S)$; then define the following quantities:
\begin{itemize}
\item $L(I,S)$: the minimum number of bb-pairings in a wiring;
\item $P(I,S,j)$: the probability that a wiring with exactly $j$ bb-pairings occurs;
\item $N(I,S,j)$: the number of all possible ways to select $j$ bb-pairings;
\item $N_w(I,S,j)$: the number of all possible wirings with a given list of $j$ bb-pairings when a (generic) clean mobile gets paired;
\item $N(I,S,j,w_t)$: the number of all possible wirings with a given list of $j$ bb-pairings and where the clean mobile $w_t$ is paired;
\item $W_w(I,S,j)=N(I,S,j)\cdot N_w(I,S,j)$: the number of all possible wirings with $j$ bb-pairings when a (generic) clean mobile gets paired;
\item $W(I,S,j,w_t)=N(I,S,j)\cdot N(I,S,j,w_t)$: the number of all possible wirings with $j$ bb-pairings where the clean mobile $w_t$ is paired;
\item $N_{\dag}(I,S,h)$: in the $\dag$ case, with $j\geq 2$, the number of possible wirings with $b_h$ unpaired, for $h \leq I$.
\end{itemize}

In the above notations, the (non recursive) closed form expression equivalent to Eq.~\ref{eq:recursive} for the probability $P(I,S)$ of a given susceptible mobile $w_t$ to get infected in a given initial configuration $(I,S)$ can be written as follows:
\begin{equation}
\boxed{
\begin{split}
P(I,S) &= \sum_{j=H(I-S-1)\frac{I-S-(I+S)\bmod 2}{2}}^{\left\lfloor \frac{I}{2} \right\rfloor} \Bigg[ H\left( \left\lfloor \frac{I}{2} \right\rfloor -j\right) H\left( j-\frac{I-S-(I+S)\bmod2}{2} \right) \cdot \\
&\phantom{=}\cdot \displaystyle{\prod_{k=0}^{I-j-1-H(I-S-1)\delta((I+S+1) \bmod 2)\delta(2j-I+S+1)} \frac{1}{I+S-1-2k}} \Bigg] \cdot\\
&\phantom{=}\cdot \displaystyle{\Bigg[  (1-H(I-S-1)\delta((I+S+1) \bmod 2)\delta(2j-I+S+1)) \frac{\displaystyle{\prod_{k=1}^{j} \binom{I-2(k-1)}{2}}}{j!}  +}\\
&\phantom{=\cdot} + \displaystyle{H(I-S-1)\delta((I+S+1) \bmod 2)\delta(2j-I+S+1)} \cdot \\
&\phantom{=\cdot} \cdot\left\{ \delta(j)+\delta(j-1)\left(\binom{I}{2}-1\right) +\right. \\
&\phantom{=\cdot} + H(j-2) \sum_{h=S+1}^I \left[\!\!\left[ \delta(h-I)\displaystyle{\frac{1}{j!}\prod_{k=0}^{j-1} \binom{I-1-2k}{2}}+ \right.\right. \\  
&\phantom{=\cdot+} +\delta(h-I+1)\displaystyle{\frac{1}{(j-1)!} (I-2)\prod_{k=0}^{j-2} \binom{I-3-2k}{2}}+ \\
&\phantom{=\cdot++} +H(h-S-1)H(I-2-h)\left( \delta(j-I+h)\displaystyle{\frac{1}{(j-I+h)!}\prod_{k=1}^{I-h} (h-k)} \right. + \\
&\phantom{=\cdot++} +H(j-I+h-1)\displaystyle{\frac{1}{(j-I+h)!}}\cdot \\
&\phantom{=\cdot+++} \cdot \left. \left. \left. \left.\left.  \displaystyle{\prod_{k=1}^{I-h} (h-k) \prod_{d=1}^{j-I+h} \binom{2h-I-1-2(d-1)}{2}}\right)\right) \right]\!\!\right] \right\} \Bigg]\cdot  \\  
&\phantom{=}\cdot \displaystyle{\Bigg[ (I-2j)!\binom{S-1}{I-1-2j}(1-H(I-S-1)\delta((I+S+1) \bmod 2)\delta(2j-I+S+1))+}\\ 
&\phantom{=} \displaystyle{\phantom{\binom{1}{1}}+S!H(I-S-1)\delta((I+S+1) \bmod 2)\delta(2j-I+S+1)\Bigg]}\ .
\end{split}
}
\label{eq:closed_form}
\end{equation}

Eq.~\ref{eq:closed_form} has its roots on the following counting argument: the probability that a given clean mobile $w_t$ gets infected is the sum over all admissible values of $j$ of all possible wirings with $j$ bb-pairings weighted by the probability that a wiring with exactly $j$ bb-pairings occurs:
\begin{equation}
\begin{split}
P(I,S) &= \sum_{j=L(I,S)}^{\left\lfloor \frac{I}{2} \right\rfloor} P(I,S,j) \cdot W(I,S,j,w_t)\\
&= \sum_{j=L(I,S)}^{\left\lfloor \frac{I}{2} \right\rfloor} P(I,S,j) \cdot N(I,S,j)\cdot N(I,S,j,w_t)\ ,
\end{split}
\label{eq:closed_form_initial}
\end{equation}
where $L(I,S)$ is the minimum number of $bb$-pairings that can be established in an initial configuration $(I,S)$.

The rationale of summing over the number of $bb$-pairings to compute $P(I,S)$ relies on the observation that the probability of $w_t$ of getting infected depends on the number of available infected mobiles that will pair with clean mobiles, that is exactly the number of infected mobiles which are not already paired to another infected mobile, \textit{i.e.}, that are not involved in a  bb-pairing.

In particular, the three terms between brackets in Eq.~\ref{eq:closed_form} match respectively the three factors in Eq.~\ref{eq:closed_form_initial}, while the term between double brackets ($[\![,]\!]$ to enhance readability) corresponds to $N_{\dag}(I,S,h)$.

In what follows we will show that the expansion of the right-hand member of Eq.~\ref{eq:closed_form_initial} coincides with Eq~\ref{eq:closed_form}.
The expansions of all terms will be carried out first by separately considering all occurring cases, and then providing an unique closed form formula (without conditional expressions) by using the Heaviside step and the Kronecker delta functions.

\begin{lemma}
Given an initial configuration $(I,S)$, the minimum number $L(I,S)$ of bb-pairings in a wiring is the following:
\begin{displaymath}
\begin{split}
L(I,S) &= 
\begin{cases}
0 & \emph{for $I\leq S$}\\
\frac{I-S}{2} & \emph{for $I>S$, $I-S\in 2\mathbb{Z}$}\\
\frac{I-S-1}{2} & \emph{for $I>S$, $I-S\in 2\mathbb{Z}+1$}
\end{cases} \\
&= H(I-S-1)\frac{I-S-(I+S)\bmod 2}{2}\ ,
\end{split}
\end{displaymath}
while the maximum number is $\left\lfloor \frac{I}{2} \right\rfloor$.
\end{lemma}
In fact, while when $I\leq S$ it is possible not to have any bb-pairing, when $I>S$ they cannot be less than $\frac{I-S}{2}$ or $\frac{I-S-1}{2}$ respectively when $I-S$ is even or odd. 
This is due to the constraint \#1 imposing that an infected mobile $b_t$ must connect to another device whenever available, when it is its turn to choose. \qed

\begin{lemma}
Given a $(I,S)$ configuration, the probability $P(I,S,j)$ that a wiring with exactly $j\geq 0$ bb-pairings between two infected mobiles occurs is the following:
\begin{displaymath}
\begin{split}
P(I,S,j) &= 
\begin{cases}
0 & \emph{if $j>\left\lfloor\frac{I}{2}\right\rfloor$ or} \\
& \emph{if $j<\frac{I-S}{2}$ when $I>S$ and $I+S\in 2\mathbb{Z}$ or} \\
& \emph{if $j<\frac{I-S-1}{2}$ when $I>S$ and $I+S\in 2\mathbb{Z}+1$} \\
& \\
\displaystyle{\prod_{k=0}^{I-j-1-z} \frac{1}{I+S-1-2k}} & \emph{otherwise,}\\
& \emph{with $z=1$ in the case $\dag$ and $0$ elsewhere} \end{cases}  \\
&= \sum_{j=H(I-S-1)\frac{I-S-(I+S)\bmod 2}{2}}^{\left\lfloor \frac{I}{2} \right\rfloor} H\left( \left\lfloor \frac{I}{2} \right\rfloor -j\right) H\left( j-\frac{I-S-(I+S)\bmod2}{2} \right) \cdot \\
&\phantom{=}\cdot \displaystyle{\prod_{k=0}^{I-j-1-H(I-S-1)\delta((I+S+1) \bmod 2)\delta(2j-I+S+1)} \frac{1}{I+S-1-2k}}\ .
\end{split}
\end{displaymath}
\end{lemma}
In fact, when there are $j$ bb-pairings in the admissible range, all possible wirings depend on the choice of $j$ infected devices $b$ and $I-2j$ clean devices $w$, \textit{i.e.} $I-j$ elements from the original sets of $I+S$. 
The first element has probability $\frac{1}{I+S-1}$ to be chosen, the second $\frac{1}{I+S-3}$, the third $\frac{1}{I+S-5}$ and so on. \qed

\begin{lemma}
Given an initial configuration $(I,S)$ in the $\dag$ case with $j\geq 2$, then the number $N_{\dag}(I,S,h)$ of possible wirings with $b_h$ unpaired, for $h \leq I$, is:
\begin{displaymath}
\begin{split}
N_{\dag}(I,S,h) &= 
\begin{cases}
\displaystyle{\frac{1}{|P(j)|} \prod_{k=0}^{j-1} |C(I-1-2k,2)}| & \emph{for $h=I$}\\
\\
\displaystyle{\frac{1}{|P(j-1)|} (I-2)\prod_{k=0}^{j-2} |C(I-3-2k,2)|} & \emph{for $h=I-1$}\\
\\
0 & \emph{for $S+1\leq h\leq I-2$}\\
& \emph{and $j<I-h$}\\
\\
\displaystyle{\frac{1}{|P(j-I+h)|}\prod_{t=1}^{I-h} (h-t)} & \emph{for $S+1\leq h\leq I-2$}\\
& \emph{and $j=I-h$}\\
\\
\displaystyle{\frac{1}{|P(j-I+h)|}\prod_{t=1}^{I-h} (h-t) \prod_{t=1}^{j-I+h} |C(2h-I-1-2(t-1),2)|} & \emph{for $S+1\leq h\leq I-2$}\\
& \emph{and $j>I-h$}\\
\\
0 & \emph{for $h\leq S$}
\end{cases}\\
&= \delta(h-I)\displaystyle{\frac{1}{j|}\prod_{k=0}^{j-1} \binom{I-1-2k}{2}} + \\
&\phantom{=} +\delta(h-I+1)\displaystyle{\frac{1}{(j-1)!} (I-2)\prod_{k=0}^{j-2} \binom{I-3-2k}{2}} +\\
&\phantom{=} +H(h-S-1)H(I-2-h) \left( \delta(j-I+h)\displaystyle{\frac{1}{(j-I+h)!}\prod_{k=1}^{I-h} (h-k)}\right. \\
&\phantom{=+} + \left. H(j-I+h-1)\displaystyle{\frac{1}{(j-I+h)!}\prod_{k=1}^{I-h} (h-k) \prod_{d=1}^{j-I+h} \binom{2h-I-1-2(d-1)}{2}} \right) \ .
\end{split}
\end{displaymath}
\end{lemma}
The idea is that all the $I-h$ infected mobiles $b_{h+1},\ldots,b_I$ must be part of a bb-pairing, so they must be connected to one of the $b_1,\ldots,b_{h-1}$. 
Once they have been chosen, the remaining $j-(I-h)$ bb-pairings must be selected among the mobiles $b_1,\ldots,b_{h-1}$ that are yet unpaired. 
Both considerations can be exploited in terms of combinations using the definitions and the properties of Fig.~\ref{fig:combinatorics}.\qed

\begin{lemma}
In the $(I,S)$ configuration, the number of all possible ways to select $j$ bb-pairings is:
\begin{displaymath}
\begin{split}
N(I,S,j) &= 
\begin{cases}
\displaystyle{\frac{\displaystyle{\prod_{k=1}^{j} |C(I-2(k-1),2)|}}{|P(j)|}}  & \emph{apart from the $\dag$ case}\\
1 & \emph{in the $\dag$ case with $j=0$} \\
|C(I,2)|-1 & \emph{in the $\dag$ case with $j=1$} \\
\displaystyle{\sum_{h=S+1}^{I} N_{\dag}(I,S,h)} & \emph{in the $\dag$ case with $j\geq 2$}
\end{cases}\\
&= \displaystyle{(1-H(I-S-1)\delta((I+S+1) \bmod 2)\delta(2j-I+S+1)) \frac{\displaystyle{\prod_{k=1}^{j} \binom{I-2(k-1)}{2}}}{j!}  +}\\
&\phantom{=\cdot} + \displaystyle{H(I-S-1)\delta((I+S+1) \bmod 2)\delta(2j-I+S+1) \left( \delta(j)+\delta(j-1)\left(\binom{I}{2}-1\right) +\right.} \\
&\phantom{=\cdot} + \left. H(j-2) \sum_{h=S+1}^I N_{\dag}(I,S,h) \right) \ .
\end{split}
\end{displaymath}
\end{lemma}
Apart from the $\dag$ case, selecting $j$ bb-pairings is equivalent to consecutively choosing $j$ unordered pairs $b_r|b_s$ from the original set of $I$ infected mobiles. 
The first pair can be chosen in $|C(I,2)|$ ways, the second pair in $|C(I-2,2)|$ and so on. 
The division by $|P(j)|$ is motivated by the fact that the particular ordering in which the $j$ pairs are chosen is irrelevant: the list $b_1|b_2, b_3|b_4, b_5|b_6$ is undistinguishable from the list $b_5|b_6, b_1|b_2, b_3|b_4$. 
The number of these different ordering is precisely $|P(j)|$ by definition of permutations.
In the $\dag$ case, if $j=0$ there is only one way to choose $0$ bb-pairings, while if $j=1$ the unpaired infected mobile can only be $b_I$, so from $|C(I,2)|$ we have to subtract the case where the only bb-pairing involves $b_I$, which is impossible.
Finally, in the $\dag$ case with $j\geq 2$ the unpaired infected mobile can be any $b_h$ with $S+1\leq h\leq I$, and the total number of cases (which coincides with the number of cases where $b_t$ is selected, since all the clean mobiles are connected in these situations) is the sum of all cases with $h=S+1,\ldots,I$. \qed

\begin{lemma}
In the $(I,S)$ configuration, with $j$ bb-pairings, the number of all possible cases when a particular $w_t$  is chosen is:
\begin{displaymath}
\begin{split}
N(I,S,j,w_t) &= 
\begin{cases}
|P(S)|  & \emph{in the $\dag$ case} \\
|P(I-2j)|\cdot|C(S-1,I-2j-1)| & \emph{otherwise\ .}
\end{cases}\\
&= \displaystyle{\left( (I-2j)!\binom{S-1}{I-1-2j}(1-H(I-S-1)\delta((I+S+1) \bmod 2)\delta(2j-I+S+1))\right. +}\\ 
&\phantom{=} \displaystyle{\left.\phantom{\binom{1}{1}}+S!H(I-S-1)\delta((I+S+1) \bmod 2)\delta(2j-I+S+1)\right)}\ .
\end{split}
\end{displaymath}
\end{lemma}
The result follows immediately from the cardinality equations in Fig.~\ref{fig:combinatorics}, in particular from the fact that among all combinations of $M$ objects in groups of $T$ elements, a particular element is selected exactly $|C(M-1,T-1)|$ times.
When $I$ is even and $j=\frac{I}{2}$ we follow the convention $\binom{A}{B}=0$ for $A>0, B<0$.
In case $\dag$, since all the non infected mobiles are selected, the possible ways to select them are exactly their permutations. \qed

This completes the expansion of Eq.~\ref{eq:closed_form_initial} into Eq.~\ref{eq:closed_form}.

Equivalence between the recursive and the closed formula can be proven by showing that Eq.~\ref{eq:closed_form} satisfies the recursive relations of Eq.~\ref{eq:recursive}. 
The analytical proof of the equivalence involves working out a large number of cumbersome identities of binomial coefficients and factorials: in the last Section, we will briefly outline a sketch of the proof in the simple case $I=S\in2\mathbb{Z}$.
Numerically, the differences between the two formul\ae\ are below machine precision for $1\leq I,S\leq 50$.

We conclude the Section with the observation that the sum of the total number of cases weighted by their corresponding probabilities adds up correctly to one:
\begin{displaymath}
\sum_{j=L(I,S)}^{\left\lfloor \frac{I}{2} \right\rfloor} P(I,S,j) \cdot W_w(I,S,j)  = \sum_{j=L(I,S)}^{\left\lfloor \frac{I}{2} \right\rfloor} P(I,S,j) \cdot N(I,S,j)\cdot N_w(I,S,j)=1 \ ,
\end{displaymath}
because of the following counting lemma.
\begin{lemma}
In the $(I,S)$ configuration with $j$ bb-pairings, the number $N_w(I,S,j)$ of all possible ways to select the remaining clean mobiles for pairing is:
\begin{equation*}
N_w(I,S,j) = 
\begin{cases}
|P(S)|  & \emph{in the $\dag$ case} \\
|D(S,I-2j)|  & \emph{otherwise\ .} 
\end{cases}
\end{equation*}
\end{lemma}
Apart from the $\dag$ case, when there are $j$ bb-pairings, $I-2j$ infected mobiles remain to be connected with $I-2j$ clean devices. 
This is equivalent to compute the number of possible sets of $I-2j$ elements from an initial set of $S$ clean mobiles: since here the ordering matters, this is the definition of dispositions (see Fig.~\ref{fig:combinatorics}) of $I-2j$ elements from an original set of $S$.  \qed

Note that, since in the case $\dag$ all the clean mobiles are selected, the two quantities $N_w(I,S,j)$ and $N(I,S,j,w_t)$ coincide.

\section*{Analytical and computational notes}
\begin{figure}[!htp]
\begin{center}
\includegraphics[width=0.8\textwidth]{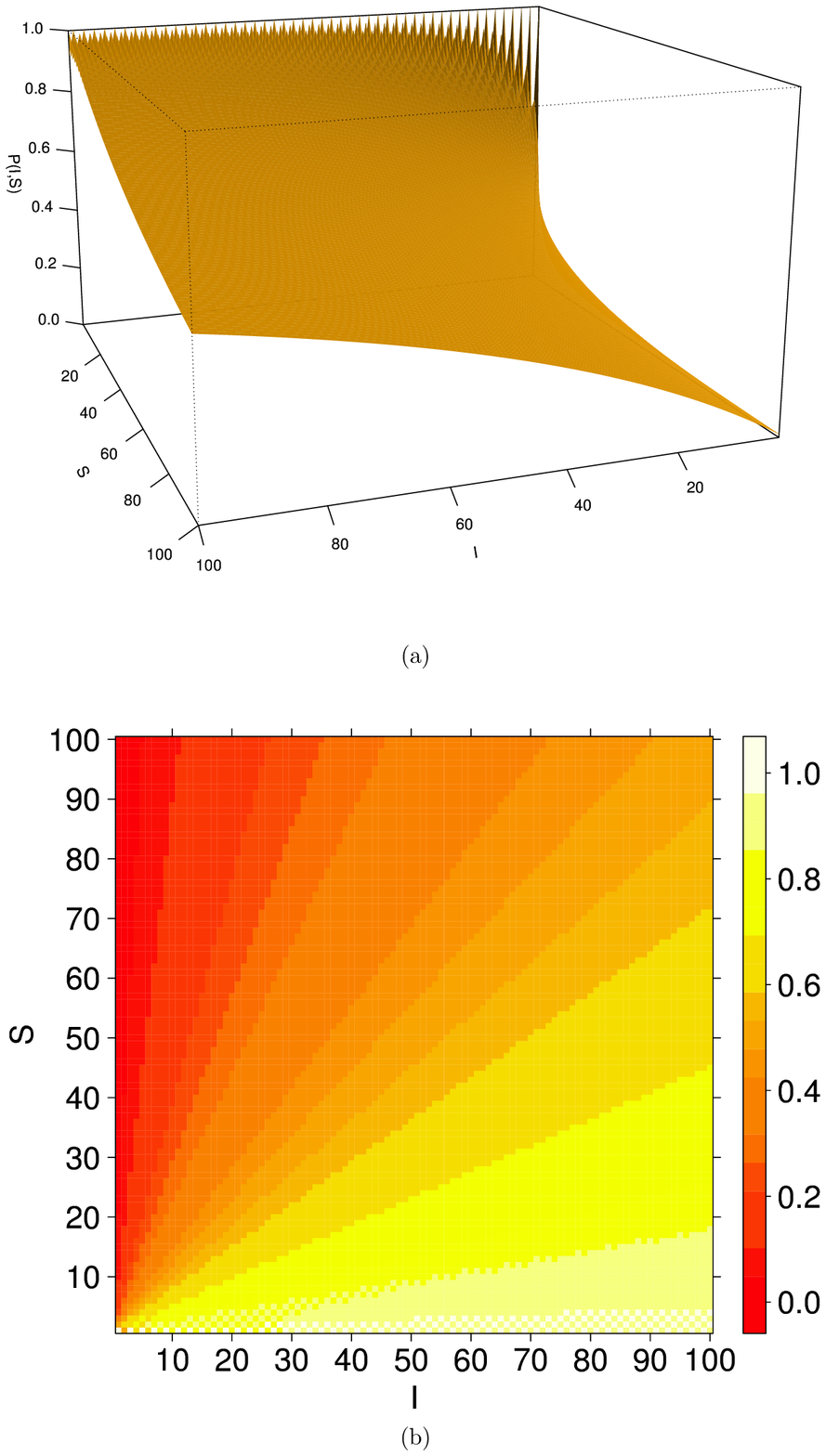}
\end{center}
\caption{\textbf{Tridimensional surface (a) and corresponding levelplot (b) of $\mathbf{P(I,S)}$ for $\mathbf{1\leq I,S\leq 100}$, linearly interpolated on the real non integer values.}}
\label{fig:surface_levelplot}
\end{figure}

Although defined only for positive integer values of $I$ and $S$, it is possible to provide a graphical sketch of the shape of the function $P(I,S)$ by linear interpolation on the non integer real values.
In Fig.~\ref{fig:surface_levelplot} we show both the tridimensional surface of $P(I,S)$ and its corresponding contourplot for values of $I$ and $S$ ranging between 1 and 100.

\begin{figure}[!htp]
\begin{center}
\includegraphics[width=0.9\textwidth]{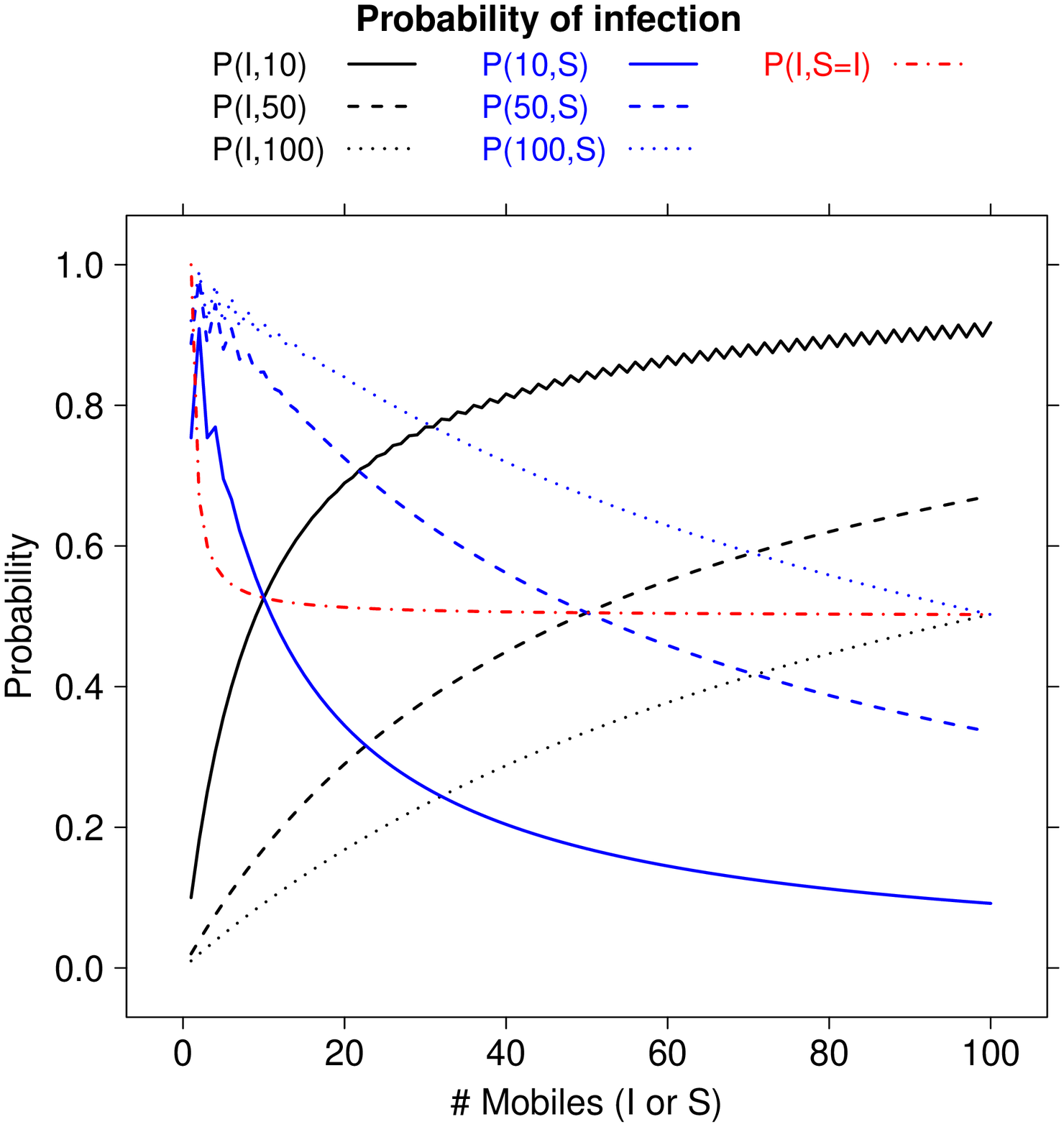}
\end{center}
\caption{
\textbf{Plot of curves of $\mathbf{P(I,S)}$ for different configurations $\mathbf{(I,S)}$.} 
In blue, we show three curves of $P(I,S)$ for constant $I$ ($I=10$ solid line, $I=50$ dashed line and $I=100$ dotted line) and $S$ ranging from 0 to 100. All three curves approach the asymptotic value 0 for increasing $S$, more rapidly for smaller values of $I$. 
In black, we show the symmetric cases obtained keeping $S$ constant ($S=10$ solid line, $S=50$ dashed line and $S=100$ dotted line) and letting $I$ range from 0 to 100. Again, all three curves approach the asymptotic value 1 for increasing $I$, more rapidly for smaller values of $S$.
The sawtooth shape of the curve $P(I,10)$ for $I\geq 30$ is due to the effect of the $\dag$ case, which induces abrupt differences in $P(I,S)$ for consecutive values of $I$ (changing from even to odd).
Finally, the dotted-dashed red line shows the curve of $P(I,S)$ for $I=S$ ranging between 0 and 100: in this case, the curve gets very close to its asymptotic value 0.5 even with small values of $I=S$; for instance, $P(10,10)\simeq 0.52$ and $P(25,25)\simeq 0.51$.
} 
\label{fig:slices}
\end{figure}

Asymptotically, the function $P(I,S)$ converges to the following limits:
\begin{equation}
\label{eq:limits}
\lim_{I\to\infty} P(I,S) =1\qquad \lim_{S\to\infty} P(I,S) = 0\qquad\lim_{\begin{smallmatrix}I,S\to\infty\\ I=S\end{smallmatrix}} P(I,S) = \frac{1}{2}\ .
\end{equation}
Graphical examples of the behaviour stated in Eq.~\ref{eq:limits} are provided in Fig.~\ref{fig:slices}, where a few curves of $P(I,S)$ are plotted when one of the two parameters is kept constant (and equal to 10, 50, 100) and the other ranges between 0 and 100, together with the curve corresponding to $P(I,S)$ for $1\leq I=S\leq 100$. When one of the two parameter is equal to a constant $T$, the smaller is $T$, the faster $P(I,S)$ converges to the limits in Eq.~\ref{eq:limits}.

\begin{figure}[!htp]
\begin{center}
\includegraphics[width=0.9\textwidth]{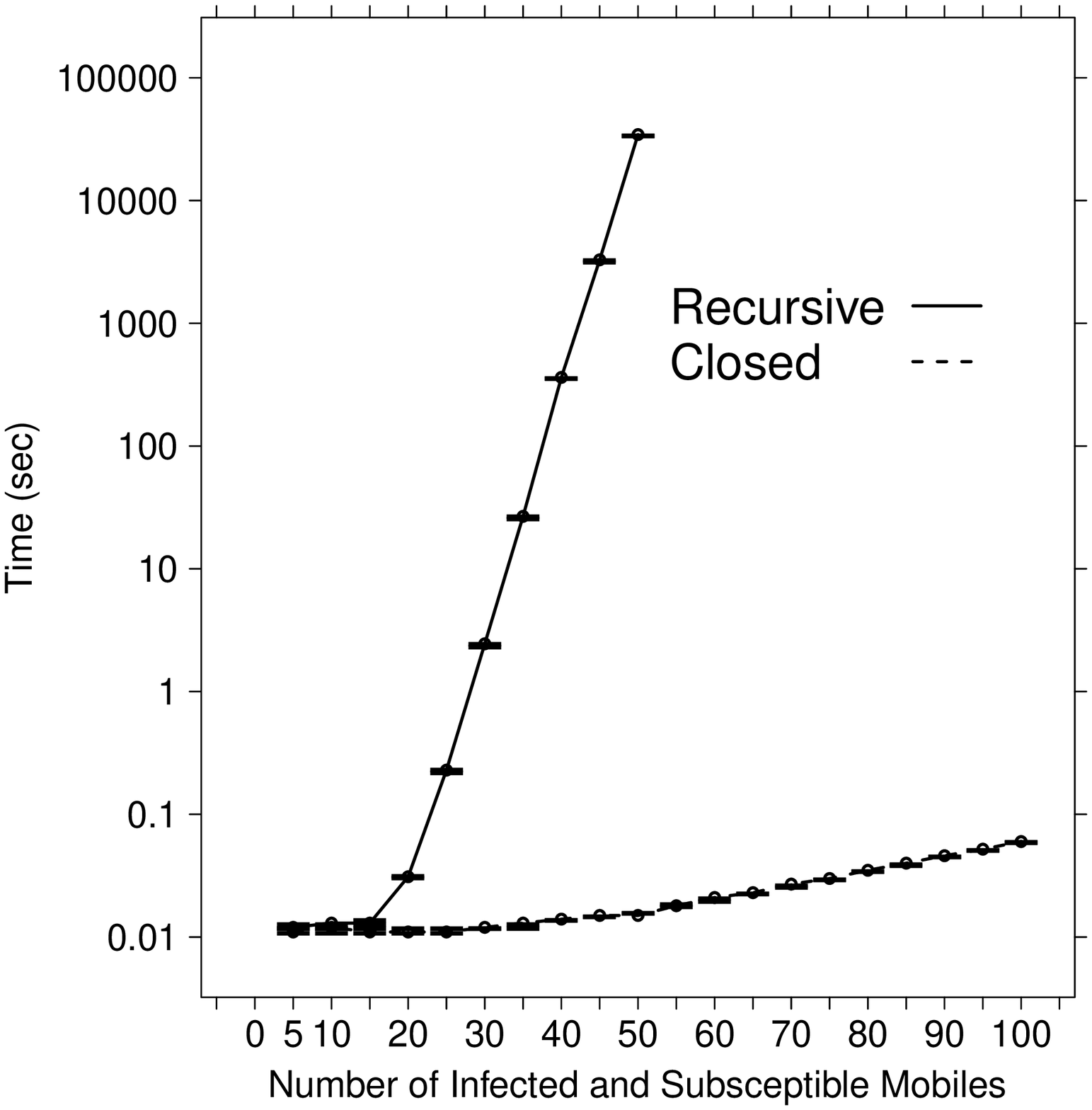}
\end{center}
\caption{
\textbf{Plot of the computing times} (in $\log$ scale) needed to compute $P(I,S)$ for different values of $I=S$ as listed in Tab.~\ref{tab:times}. Error bars range between minimum and maximum, while lines connect mean values; all values refer to 10 replicates. 
Solid line represents computing times obtained by using the recursive formula Eq.~\ref{eq:recursive}, while dotted line corresponds to the values produced by using the closed formula Eq.~\ref{eq:closed_form}.
} 
\label{fig:times}
\end{figure}

Apart from its intrinsic theoretical relevance, the non recursive closed formula is essential for numerically compute $P(I,S)$.
In fact, the computational cost is notably different by using either the recursive formula Eq.~\ref{fig:recursive} or its closed form counterpart Eq.~\ref{eq:closed_form}: namely, the explicit formula is much faster, as shown by the values reported in Tab.~\ref{tab:times} and the curves plotted in Fig.~\ref{fig:times}. 

\begin{table}[!htp]
\caption{
\textbf{Computing times (in seconds) required to compute $\mathbf{P(I,S)}$} by the recursive formula in Eq.~\ref{eq:recursive} and the equivalent closed formula in Eq.~\ref{eq:closed_form}, for different values of the number of infected (I) and susceptible (S). 
In particular, $I=S=5\ldots 100$, and only the closed formula was used for $I,S>50$ (due to the excessively long runtimes: \textit{e.g.}, computing $P(50,50)$ by the recursive formula took more than 9 hours). 
Mean, maximum (Max) and minimum (Min)  values for 10 replicates of each experiment are reported. 
All simulations were run on a 24 core Intel Xeon E5649 CPU 2.53GHz workstation with 47 GB RAM, Linux 2.6.32 (Red Hat 4.4.6), with software written in Python 2.6.6.
}
\begin{center}
\begin{tabular}{rrrrrrr}
\hline
I=S & \multicolumn{3}{c}{Recursive} &  \multicolumn{3}{c}{Closed Form} \\
& Min & Mean & Max & Min & Mean & Max \\ 
\hline
5 & 0.012 & 0.012 & 0.013 & 0.011 & 0.011 & 0.012 \\ 
10 & 0.012 & 0.013 & 0.013 & 0.011 & 0.012 & 0.012 \\ 
15 & 0.013 & 0.013 & 0.014 & 0.011 & 0.011 & 0.012 \\ 
20 & 0.031 & 0.031 & 0.032 & 0.011 & 0.011 & 0.012 \\ 
25 & 0.223 & 0.229 & 0.235 & 0.011 & 0.011 & 0.012 \\ 
30 & 2.365 & 2.449 & 2.491 & 0.012 & 0.012 & 0.012 \\ 
35 & 26.203 & 26.757 & 27.419 & 0.012 & 0.013 & 0.013 \\ 
40 & 361.621 & 362.351 & 362.894 & 0.014 & 0.014 & 0.014 \\ 
45 & 3225.718 & 3287.492 & 3333.242 & 0.015 & 0.015 & 0.015 \\ 
50 & 34336.694 & 34433.664 & 34555.204 & 0.016 & 0.015 & 0.016 \\ 
55 &  &  &  & 0.018 & 0.018 & 0.019 \\ 
60 &  &  &  & 0.020 & 0.021 & 0.021 \\ 
65 &  &  &  & 0.023 & 0.023 & 0.023 \\ 
70 &  &  &  & 0.026 & 0.027 & 0.027 \\ 
75 &  &  &  & 0.030 & 0.030 & 0.030 \\ 
80 &  &  &  & 0.035 & 0.035 & 0.035 \\ 
85 &  &  &  & 0.039 & 0.040 & 0.040 \\ 
90 &  &  &  & 0.046 & 0.046 & 0.046 \\ 
95 &  &  &  & 0.052 & 0.052 & 0.052 \\ 
100 &  &  &  & 0.060 & 0.060 & 0.061 \\ 
\hline
\end{tabular}
\end{center}
\begin{flushleft}
\end{flushleft}
\label{tab:times}
\end{table}

For the recursive formula the computing time shows an exponentially growing trends for increasing values of $I$ and $S$, while for the non recursive formula the computing time is very small and minimally growing for $I$ and $S$ ranging between 0 and 100.
Actually, the average time over 10 values using a Python implementation of the non recursive formula on a 24 core Intel Xeon E5649 CPU 2.53GHz Linux workstation with 47 GB RAM is 11 milliseconds for $I=S=5$ and 60 milliseconds for $I=S=10$, with very limited standard deviation.
On the same hardware, a Python implementation of the recursive formula took about 12 milliseconds for $P(5,5)$, 2.4 seconds for $P(30,30)$, 6 minutes for $P(40,40)$ and more than 9 hours for $P(50,50)$, which was the largest tested value.

\section*{Proof of equivalence in the case $\mathbf{I=S\in 2\mathbb{Z}}$}
In this Section we show the kind of arguments involved in proving the equivalence between Eq.~\ref{eq:recursive} and Eq.~\ref{eq:closed_form} by outlining the main steps of the proof in a simple case, \textit{i.e.}, when there as many infected as clean mobiles, and their numnber is even.
Clearly, the general case is computationally far more complex, but it used the same ideas.

Proving the equivalence between the recursive and the combinatorial formula requires substituting the explicit expression for $P(I,S)$ of Eq.~\ref{eq:closed_form} in its three occurrences in Eq.~\ref{eq:recursive}.
We are assuming $I=S=2x\in 2\mathbb{Z}$, thus in this case the identity we need to prove reads as follows:
\begin{displaymath}
P(2x,2x) = \frac{1}{4x-1} +\frac{2x-1}{4x-1} P(2x-1,2x-1) + \frac{2x-1}{4x-1} P(2x-2,2x)\ ,
\end{displaymath}
or, equivalently:
\begin{equation}
U(x)=(4x-1)P(2x,2x) - (2x-1) [ P(2x-1,2x-1) + P(2x-2,2x)] =1\ .
\label{eq:proof}
\end{equation}
The expression for $P(I,S)$ becomes:
\begin{displaymath}
\begin{split}
P(2x,2x) &= \sum_{j=0}^{x-1} \prod_{k=0}^{2x-j-1} \frac{1}{4x-1-2k} \frac{1}{j!} \prod_{k=1}^j \binom{2x-2k+2}{2} (2x-2j)! \binom{2x-1}{2x-2j-1} \\
&= \sum_{j=0}^{x-1}  \prod_{k=0}^{2x-j-1} \frac{1}{4x-1-2k}  \frac{1}{j!}  \prod_{k=1}^j (x-k-1) \prod_{k=1}^j (2x-2k+1) \frac{(2x-2j)!(2x-1)!}{(2x-1-2j)!(2j)!}\\
&= \sum_{j=0}^{x-1} \frac{(2j-1)!!}{(4x-1)!!} \frac{1}{j!} \frac{x!}{(x-j)!} \frac{(2x-1)!!}{(2x-2j-1)!!} (2x-2j) \frac{(2x-1)!}{(2j)!}\ ,
\end{split}
\end{displaymath}
where the upper bound is $x-1$ since the right-hand member vanishes for $j=x$ and the product symbols were eliminated by using the factorial and double factorial notations:
\begin{displaymath}
\prod_{i=a}^b i = \frac{b!}{(a-1)!}\qquad\qquad 
n!! = 
\begin{cases}
1 & \textrm{ for $n=0,-1$} \\
\displaystyle{\prod_{j=0}^{\frac{n-1}{2}} (n-2k)} = n\cdot (n-2)\cdot (n-4)\cdots 3\cdot 1 & \textrm{for $n\in 2\mathbb{Z}_{>0} -1$} \\
\displaystyle{\prod_{j=0}^{\frac{n-2}{2}} (n-2k)} = n\cdot (n-2)\cdot (n-4)\cdots 4\cdot 2 & \textrm{for $n\in 2\mathbb{Z}_{>0}$} \ .
\end{cases}
\end{displaymath}
Analogously, the expansions for $P(I-1,S-1)$ and $P(I-2,S)$ become respectively:
\begin{displaymath}
\begin{split}
P(2x-1,2x-1) &= \sum_{j=0}^{x-1} \frac{(2j-1)!!}{(4x-3)!!} \frac{1}{j!} \frac{(x-1)!}{(x-j-1)!} \frac{(2x-1)!!}{(2x-2j-1)!!} (2x-2j-1) \frac{(2x-2)!}{(2j)!} \\
P(2x-2,2x) &= \sum_{j=0}^{x-1} \frac{(2j+1)!!}{(4x-3)!!} \frac{1}{j!} \frac{(x-1)!}{(x-j-1)!} \frac{(2x-3)!!}{(2x-2j-3)!!} (2x-2j-2) \frac{(2x-1)!}{(2j+2)!}\ .
\end{split}
\end{displaymath}
Then the left-hand member of Eq.~\ref{eq:proof} reads as follows:
\begin{displaymath}
\begin{split}
U(x) = \sum_{j=0}^{x-1} \phantom{-} &\frac{(2j-1)!!(x)!(2x-1)!!(2x-2j)(2x-1)!(4x-1)}{(4x-1)!!(j)!(x-j)!(2x-2j-1)!!(2j)!} +\\
- &\frac{(2j-1)!!(x-1)!(2x-1)!!(2x-2j-1)(2x-2)!(2x-1)}{(4x-3)!!(j)!(x-j-1)!(2x-2j-1)!!(2j)!} +\\
- &\frac{(2j+1)!!(x-1)!(2x-3)!!(2x-2j-2)(2x-1)!(2x-1)}{(4x-3)!!(j)!(x-j-1)!(2x-2j-3)!!(2j+2)!} \ ,
\end{split}
\end{displaymath}
which, collecting common factors, reduces to:
\begin{displaymath}
\begin{split}
U(x) &= \sum_{j=0}^{x-1}  \frac{(2j-1)!!(x-1)!(2x-1)!!(2x-1)!}{(4x-3)!!(j)!(x-j-1)!(2x-2j-3)!!(2j)!} \left[ \frac{2x}{2x-2j-1} -1- \frac{x-j-1}{j+1}\right] \\
&= \sum_{j=0}^{x-1}  \frac{(2j-1)!!x!(2x-1)!!(2x-1)!(4j-2x+3)}{(4x-3)!!(j+1)!(x-j-1)!(2x-2j-1)!!(2j)!} \ .
\end{split}
\end{displaymath}
Now, expanding the double factorial by the identity:
\begin{displaymath}
(2n-1)!! = \frac{(2n)!}{2^n n}\ ,
\end{displaymath}
and carrying the terms not involving $j$ outside the summation symbol, the above quantity becomes:
\begin{equation}
\label{eq:almost_done}
U(x) = \frac{ 2^{2x-1} ((2x-1)!)^2 (2x)!  }{2(2x-1))!} \sum_{j=0}^{x-1} \frac{(x-j)(4j-2x+3)}{4^j (j+1) (j!)^2 (2(x-j))!}\ .
\end{equation}
Now, applying the following identity
\begin{displaymath}
\frac{(2n)!}{ 2^n (n!)^2 (n+1)!} =  \sum_{z=0}^{\frac{x-1}{2}} \frac{4z-n+2}{2^{2z+1} (z+1) (z!)^2 (n-2z)!}
\end{displaymath}
to Eq.~\ref{eq:almost_done} with $n=2x-1$, we obtain that
\begin{displaymath}
U(x)=1
\end{displaymath}
as claimed. \qed

\section*{Acknowledgments}
The authors thank two anonymous referees for their precious suggestions and notes, which helped in greatly improving the paper.

\bibliography{merler12combinatorial}
\end{document}